\documentclass[preprint,12pt,longnamesfirst]{aastex}

\usepackage{latexsym}
\shorttitle{X-rays from WR in the MCs}
\shortauthors{Guerrero et al.}

\begin{document}

\title{An X-ray Survey of Wolf-Rayet Stars in the Magellanic Clouds. II. 
       The ROSAT PSPC and HRI Datasets}

\author{Mart\'{\i}n A.\ Guerrero\altaffilmark{1,2}, 
        You-Hua Chu\altaffilmark{2}} 
\altaffiltext{1}{Instituto de Astrof\'{\i}sica de Andaluc\'{\i}a, CSIC, 
        Apdo.\ 3004, 18080, Granada, Spain}
\altaffiltext{2}{Astronomy Department, University of Illinois, 
        1002 W. Green Street, Urbana, IL 61801, USA}

\begin{abstract}

Wolf-Rayet (WR) stars in the Magellanic Clouds (MCs) are ideal for 
studying the production of X-ray emission by their strong fast stellar 
winds.  
We have started a systematic survey for X-ray emission from WR stars in 
the MCs using archival \emph{Chandra}, \emph{ROSAT}, and \emph{XMM-Newton} 
observations.  
In Paper I, we reported the detection of X-ray emission from 29 WR
stars using \emph{Chandra} ACIS observations of 70 WR stars in the MCs.  
In this paper, we report the search and analysis of archival \emph{ROSAT}
PSPC and HRI observations of WR stars.
While useful \emph{ROSAT} observations are available for 117 WR
stars in the MCs, X-ray emission is detected from only 7 of them.
The detection rate of X-ray emission from MCs WR stars in the \emph{ROSAT} 
survey is much smaller than in the \emph{Chandra} ACIS survey,
illustrating the necessity of high angular resolution and sensitivity.
LMC-WR\,101-102 and 116 were detected by both \emph{ROSAT} and 
\emph{Chandra}, but no large long-term variations are evident.

\end{abstract}


\section{Introduction}

Wolf-Rayet (WR) stars are characterized by their broad emission lines, 
indicating copious fast stellar winds.
Spectral analyses of WR stars show typical wind terminal velocities 
of 1,000--3,000 km~s$^{-1}$ \citep{PBH90} and mass loss rates of a 
few $\times10^{-5} M_\odot$~yr$^{-1}$ \citep{dJNvH88}.
Such powerful stellar winds are expected to generate a variety of
X-ray sources.
Within the WR wind itself, instability shocks produce regions with high 
temperatures and high densities for X-ray emission \citep{LW80,GO95}.
Upon leaving the WR star, the wind may encounter a massive companion's
wind, and the colliding winds produce compressed hot gas that
emits X-rays.
Finally, as the WR wind impinges on the ambient medium, a wind-blown 
bubble may form, and the shocked WR wind in the bubble interior may 
emit in X-rays \citep{Wetal77,GML96,GLM96}.
Therefore, X-ray observations of WR stars provide an opportunity to
probe the opacity of the stellar wind, study the orbital configuration
of a WR+OB binary system, and examine the stellar mechanical energy
injection into the interstellar medium.

Previous \emph{Einstein} and \emph{ROSAT} X-ray surveys of Galactic 
WR stars have shown that WR stars in binary systems have higher 
$L_{\rm X}/L_{\rm bol}$ than single WR stars or O stars, and that 
single WN stars are generally brighter than WC stars, although no 
simple $L_{\rm X}/L_{\rm bol}$ relationship appears to exist among 
single WN stars \citep{P87,PHC95,W96}. 
While these results reveal the potential of scientific yields from
X-ray observations of WR stars, the Galactic sample is plagued by
heavy obscuration in the Galactic plane that renders a large fraction
of WR stars undetectable.
The nearby Large and Small Magellanic Clouds (LMC \& SMC) are ideal 
locations to expand the X-ray observations of WR stars, since the 
foreground and internal extinctions in these galaxies are small.  
Furthermore, their lower metallicities allow us to probe abundance
effects on the stellar winds, and their known distances allow us to
determine $L_{\rm X}$ with greater certainty.

In the first paper of this series \citep[][hereafter Paper I]{GC07}, 
we have used the current archive of the \emph{Chandra X-ray 
Observatory} to search for X-ray emission from WR stars in the 
Magellanic Clouds (MCs).  
This survey included useful ACIS observations for 70 of the 146 
known WR stars in the MCs and resulted in credible detection of 
X-ray emission from 29 of these WR stars and possibly another 4
WR stars.
Many of the WR stars in the MCs that have not been observed by 
\emph{Chandra} have \emph{ROSAT} X-ray observations available 
in the archive.  
In this paper, we have used the entire \emph{ROSAT} archive of pointed 
observations to search for and analyze X-ray sources associated 
with WR stars in the MCs in order to complement our \emph{Chandra} 
ACIS X-ray survey for WR stars in the MCs and to investigate 
long-term X-ray variability.  
In an upcoming paper (Carter et al., in preparation, Paper III), 
the results from the \emph{ROSAT} and \emph{Chandra} surveys of 
WR stars in the MCs will be complemented by an \emph{XMM-Newton} 
survey.  
All three archival studies will be analyzed together in conjunction with 
a systematic spectroscopic search for binaries for all WR stars in the 
MCs \citep[][Schnurr et al., in preparation]{BMN01, FMG03a, FMG03b} to 
determine accurately the origin of X-ray emission from WR stars.

\section{ROSAT Observations of WR Stars in the MCs}

The \emph{R\"ontgen Satellite (ROSAT)} had two types of X-ray detectors 
onboard - the Position Sensitive Proportional Counter (PSPC) and the 
High-Resolution Imager (HRI).
The PSPC has an on-axis angular resolution of $\sim30''$ and a 
spectral resolution of $\sim$45\% at 1 keV;  it is sensitive in 
the energy range of 0.1--2.4 keV, and has a field-of-view of 
$\sim$2\arcdeg.  
The HRI has a higher angular resolution, $\sim$5\arcsec, but does 
not provide spectral resolution over the operational energy range 
of 0.1--2.0 keV;  its field-of-view is $\sim$38\arcmin.  
During the \emph{ROSAT} mission from 1990 to 1999, numerous pointed 
observations of targets in the MCs were made, and the large field-of-view
serendipitously included many WR stars.
These observations can be retrieved from the \emph{ROSAT} 
archive\footnote{\emph{ROSAT} archival data can be obtained from the 
anonymous ftp site legacy.gsfc.nasa.gov, or downloaded from the web site 
http://heasarc.gsfc.nasa.gov/W3Browse.} 
maintained by the High Energy Astrophysics Science Archive Research 
Center (HEASARC) of Goddard Space Flight Center, NASA.

To search for X-ray observations of WR stars in the MCs, we used 
the lists of LMC WR stars compiled by \citet{BAT99} and SMC WR 
stars compiled by \citet{MOP03}.  
We use only archival \emph{ROSAT} observations with exposure times
greater than 1 ks.
The point-spread-function (PSF) and effective exposure of the
telescope degrade significantly with distance from the field
center; therefore, we select only PSPC observations with
WR stars within the central 35\arcmin\ radius and HRI observations
with WR stars within the central 18\arcmin\ radius.
For PSPC observations with exposure times longer than 5 ks, we relax
the selection criteria to include observations with WR stars within
35\arcmin\ to 40\arcmin\ from the central pointing.
The search finds PSPC observations for 121 WR stars in 
the LMC and 11 WR stars in the SMC, and HRI observations for 110 WR 
stars in the LMC and 11 WR stars in the SMC.  
Tables 1 and 2 list, respectively, the available PSPC and HRI
observations, exposure times, and offsets of WR stars from the central
pointings.
Multiple observations with the same instrument of a WR star were merged 
in order to increase the signal-to-noise ratio.

\section{Results}

X-ray images are extracted from the merged PSPC and/or HRI observations 
of each WR star in the MCs within the full spectral energy range, i.e., 
0.1--2.4 keV for the PSPC and 0.1--2.0 keV for the HRI.  
A pixel size of 5\arcsec\ pixel$^{-1}$ is used for the PSPC images
and 2\arcsec\ pixel$^{-1}$ for the HRI images.
These images are subsequently smoothed with a Gaussian profile of FWHM 
of 15\arcsec\ for the PSPC and 3\arcsec\ for the HRI.
The smoothed images are used to search for X-ray emission at the
location of WR stars.  
When X-ray emission is detected within 30\arcsec\ from the location 
of a WR star, we compare the X-ray images of the WR star with an
optical image extracted from the Digitized Sky Survey\footnote{
The Digitized Sky Survey (DSS) is based on photographic data obtained 
using the UK Schmidt Telescope and the Oschin Schmidt Telescope on 
Palomar Mountain.  
The UK Schmidt was operate by the Royal Observatory of Edinburgh, with 
funding from the UK Science and Engineering Research Council, until 1988 
June, and thereafter by the Anglo-Australian Observatory. 
The Palomar Observatory Sky Survey was funded by the National Geographic 
Society. 
The Oschin Schmidt Telescope is operated by the California Institute of 
Technology and Palomar Observatory. 
The plates were processed into the present compressed digital form with 
the permission of these institutes. 
The Digitized Sky Survey was produced at the Space Telescope Science 
Institute under US government grant NAGW-2166.} (DSS)
to search for a point source at the location of the star or diffuse
emission from its surrounding bubble, if it exists.  
To assess the reliability of these detections, we have defined a source 
region encompassing the X-ray source at the location of the WR star 
and an appropriate background region without sources, and computed the 
background-subtracted \emph{ROSAT} PSPC and/or HRI counts within the 
source region using the 
IRAF\footnote{IRAF is distributed by the National Optical Astronomy 
Observatories, which is operated by the Association of Universities for 
Research in Astronomy, Inc. (AURA) under cooperative agreement with the 
National Science Foundation.}  
PROS task \emph{imcnts}.  
This has allowed us to confirm the $\gtrsim3\sigma$ detections of X-ray 
emission from the 7 WR stars in the LMC listed in Table~3.
The net count rates and net counts of these WR stars are listed in columns 
5 and 6 of Tab.~3, respectively.  
No WR stars in the SMC are detected by \emph{ROSAT}.
The correlation between these detections and sources in different 
\emph{ROSAT} catalogs is listed in Table~4.  
The \emph{ROSAT} PSPC and HRI X-ray and DSS optical images of the WR stars 
detected in X-rays are presented in Figures~1 and 2.  
For the central regions of R\,136 and R\,140 (Fig.~2), only the HRI 
images are shown as the angular resolution of the PSPC observations 
is too poor to resolve these stars.

The WR stars in the MCs that are not detected by \emph{ROSAT} observations 
are listed in Table~5.  
For these WR stars, we use source regions with sizes matching the PSF 
of \emph{ROSAT} PSPC and HRI in order to determine their 3-$\sigma$ 
upper limits using the IRAF PROS task \emph{imcnts}.  
The radii of these source regions range from 20\arcsec\ to
60\arcsec\ for the PSPC and from 10\arcsec\ to 20\arcsec\ 
for the HRI, depending on the offsets of the WR stars from
the central pointings.  
The resulting 3-$\sigma$ upper limits are listed in column 5 of Tab.~5.  
The distribution of these upper limits indicates that most of the 
non-detections have HRI count rates $<$1.0$\times$10$^{-4}$~cnts~s$^{-1}$ 
and PSPC count rates $<$1.5$\times$10$^{-4}$~cnts~s$^{-1}$.

Several WR stars in the MCs are found to be embedded in diffuse X-ray 
emission or close to bright X-ray sources.  
The analysis of these sources is neither possible nor necessary, since
\emph{Chandra} observations provide a much clearer view.
These include \emph{ROSAT} PSPC observations of LMC-WR\,96, 97, 98, 
99, 100, 101, 102, 103, 104, 105, 106, 107, 108, 109, 110, 111, 112,
113, 114, 115, and 116, all within 70$''$ from R\,136,
and PSPC observations of LMC-WR\,31, 81, 82, 84, 85, 88, 89, 91, 92, 
93, 95, 117, 118, 119, 121, and 122, and of HD\,5980 in the SMC, 
as they are embedded in diffuse X-ray emission.   
Similarly, no analysis was attempted for the \emph{ROSAT} HRI 
observations of LMC-WR\,99, 100, 104, 105, 106, 107, 108, 109, 
110, 111, 112, 113, 114, and 115 near R\,136, the observations 
of LMC-WR\,91 and 93 that are too close to bright X-ray sources, 
or the observations of LMC-WR\,80, 85, 92, and 118, and HD\,5980 
in the SMC, which are superposed by bright diffuse X-ray emission.

\subsection{Comparison Between the Chandra ACIS and ROSAT Surveys}

The \emph{Chandra} ACIS and \emph{ROSAT} surveys for X-ray emission 
from WR stars in the MCs have many stars in common.  
LMC-WR\,19, 20, 67, 78, 79, 119, 125, 126, and 127 are detected by 
\emph{Chandra} ACIS, but not \emph{ROSAT}.
In all these cases, the \emph{ROSAT} PSPC and HRI count rates
expected from their \emph{Chandra} ACIS count rates are below 
the 3-$\sigma$ upper limit listed in Tab.~5.   
LMC-WR\,101-102 and 116 in the LMC are detected both by \emph{Chandra} 
ACIS and by \emph{ROSAT} HRI.  
The \emph{Chandra} ACIS count rates and spectral properties of 
these two sources (Paper I) correspond to \emph{ROSAT} HRI count rates 
of 2.8$\times$10$^{-3}$ cnts~s$^{-1}$ for LMC-WR\,101-102, and 
(4.5--6.3)$\times$10$^{-3}$ cnts~s$^{-1}$ for the X-ray variable 
LMC-WR\,116. 
These values are fairly consistent with the HRI count rates of these 
sources listed in Tab.~3.

\subsection{X-ray Luminosity of the Wolf-Rayet Stars Detected by ROSAT}

The \emph{ROSAT} PSPC detections of WR stars in the MCs have yielded 
insufficient numbers of counts for spectral fits, and the 
\emph{ROSAT} HRI observations do not provide spectral information.  
In order to estimate the X-ray luminosities of the WR stars in the 
MCs detected by \emph{ROSAT}, we adopt the emission model
that describes the integrated spectra of the weakly detected WR 
stars in the \emph{Chandra} ACIS survey (Paper I), i.e., a thin 
plasma with a temperature of $kT$ = 1.6 keV absorbed by 
intervening material with abundances of 0.33 $Z_\odot$ 
and an absorption column density of 3$\times$10$^{21}$~cm$^{-2}$.  
We have used PIMMS to convert the \emph{ROSAT} PSPC and 
HRI count rates to X-ray luminosities in the 0.5-7.0 keV 
band and listed them in Tab.~3.

\subsection{Remarks on Individual Objects}

LMC-WR\,10 (Brey\,9) is at the core of the OB association LH 9 
in N11 that includes up to 25 stellar components in a field of 
view 6\farcs4$\times$6\farcs4 \citep{Setal95,Betal96}.  
Therefore, the X-ray emission reported in this paper may have 
an origin different from LMC-WR\,10.

The X-ray emission from LMC-WR\,38, 39 and 42 has been previously 
reported by \citet{DPC01}.  
LMC-WR\,38 is also identified as source \#538 by \citet{HP99}.  
LMC-WR\,38 (Brey\,31) and 39 (Brey\,32) are confirmed binary systems 
with periods of $\sim$3 and $\sim$2 days, respectively \citep{MNM90}.  
Similarly, LMC-WR\,42 (Brey\,34) is a binary system but with a 
longer period, 30 days \citep{SML91}.

The X-ray detection of LMC-WR\,47 (Brey\,39) needs to be examined 
carefully.  
LMC-WR\,47 is located on an area of diffuse X-ray emission and its 
number of counts is only $\sim$3.5$\sigma$ over the local background 
level of X-ray emission.  
Furthermore, the X-ray contours shown in Fig.~1 reveal a noticeable 
offset of $\sim$25\arcsec\ between the X-ray peak and the location 
of this WR star.  
Therefore, until new X-ray observations with better angular resolution 
of LMC-WR\,47 are acquired, its X-ray detection reported in this paper 
should be considered tentative.

LMC-WR\,101, 102, and 103 are in the visual multiple system R\,140 near the 
core of 30 Doradus, of which LMC-WR\,102 (R\,140a2) is a close spectroscopic 
binary with a period of $\sim$3 days \citep{Metal87}.  
LMC-WR\,101, 102, and 103 were marginally detected by the \emph{Einstein} 
High Resolution Imager \citep{WH91}.  
\emph{ROSAT} made a clear detection, being listed as source \#299 by 
\citet{SHP00}, but the X-ray emission of the different components was 
not individually resolved until the \emph{Chandra} ACIS-I observations 
of the 30 Doradus nebula offered a sharper view of this region 
\citep[][Paper I]{PPL02,Tetal06}.  
\emph{Chandra} observations show that LMC-WR\,101-102 (R\,140a1 and 
R\,140a2) are much brighter than R\,140b \citep[][Paper I]{PPL02,Tetal06}.  
The present analysis show that the level of X-ray emission in the 
\emph{ROSAT} HRI and \emph{Chandra} ACIS-I observations are consistent 
with each other.  
We note, however, that \citet{W95} reported a higher X-ray flux 
based on the \emph{ROSAT} HRI observation rh600228 
obtained in 1992 December and 1993 June.  
We have examined these individual observations, as well as the \emph{ROSAT} 
HRI observation rh400779 (1996 August and 1997 April), and find that
in all cases the HRI count rates remain at roughly a constant level 
consistent with the count rate of 2.9$\times$10$^{-3}$ cnts~s$^{-1}$ 
reported in Tab.~3.

LMC-WR\,116 (Brey\,84) is also a WR star in the 30 Doradus region 
that was marginally detected by the \emph{Einstein} High Resolution 
Imager \citep{WH91}.  
\citet{W95} reported a \emph{ROSAT} HRI count rate of 8.1$\times$10$^{-3}$ 
cnts~s$^{-1}$ based on the observation rh600228 obtained in 1992 December
and 1993 June.
A similar \emph{ROSAT} HRI count rate of 8.5$\times$10$^{-3}$ cnts~s$^{-1}$ 
is reported by \citet{SHP00} who assigned it the source number \#301 in their 
catalog.  
The \emph{ROSAT} HRI count rate reported in Tab.~3 is $\sim$40\% lower 
than the values reported by \citet{W95} and \citet{SHP00} because we 
used a smaller source aperture to exclude the contribution from 
LMC-WR\,112 (R\,136c), LMC-WR\,99 (Brey\,78), and an X-ray bright source 
near LMC-WR\,115 (Brey\,83) north of Brey\,84 (see Fig.~1d of Guerrero 
\& Chu 2007).  
If contributions from these bright neighboring sources are added to our
measurement, we recover the \emph{ROSAT} HRI count rates reported by 
\citet{W95} and \citet{SHP00}.  
To further investigate possible long-term variations of this 
source, we have analyzed the individual \emph{ROSAT} HRI 
observations rh500036 (1992 February), rh600228 (1992 December), 
and rh400779 (1996 August and 1997 April), and find HRI count 
rates of 
(5.9$\pm$0.8)$\times$10$^{-3}$ cnts~s$^{-1}$, 
(4.3$\pm$0.4)$\times$10$^{-3}$ cnts~s$^{-1}$, 
(4.0$\pm$0.4)$\times$10$^{-3}$ cnts~s$^{-1}$, and 
(4.9$\pm$0.3)$\times$10$^{-3}$ cnts~s$^{-1}$, respectively.  
These values are consistent with the \emph{ROSAT} HRI count rate 
of (4.5--6.3)$\times$10$^{-3}$ cnts~s$^{-1}$ expected from the 
\emph{Chandra} ACIS observation of Brey\,84; thus, there is no
evidence for large long-term variations.

\section{Summary}

We have searched the entire \emph{ROSAT} archive for pointed 
observations that serendipitously cover WR stars in the MCs.  
This search has yielded useful PSPC observations for 90 WR stars in 
the LMC and 10 WR stars in the SMC, and HRI observations for 87 
WR stars in the LMC and 10 WR stars in the SMC.  
A total of 117 WR stars in the MCs have useful \emph{ROSAT}
observations.

We have examined the \emph{ROSAT} observations of these 117 WR 
stars in the MCs and found X-ray emission from 7 of them, of 
which 5 had been previously reported to exhibit X-ray emission.  
We find that the X-ray detection of LMC-WR\,10 (Brey\,9) and LMC-WR\,47 
(Brey\,39) need to be confirmed by X-ray observations at higher angular 
resolution.  
The detection rate, $\sim$6\%, is much lower than that of the 
\emph{Chandra} ACIS survey, 40--50\%.  
This illustrates that the sensitivity and angular resolution 
of \emph{Chandra} is needed to study WR stars in the MCs.
Indeed, many WR stars detected by \emph{Chandra} have 
X-ray emission at levels below the 3-$\sigma$ upper limits of the 
available \emph{ROSAT} observations, are located near bright 
X-ray sources, or are superposed on bright diffuse X-ray emission,
making it difficult for \emph{ROSAT} to detect them.
Together, the \emph{ROSAT} and \emph{Chandra} surveys have 
detected X-ray emission from 34 WR stars in the MCs.

\acknowledgments

This work is supported by the \emph{Chandra X-ray Observatory} 
grant AR3-4001X.  
M.A.G.\ also acknowledges support from the grants AYA 2002-00376
and AYA 2005-01495 of the Spanish MEC (co-funded by FEDER funds)
and the Spanish program Ram\'on y Cajal.

\clearpage

{\tiny
\begin{deluxetable}{llrrr|llrrr|llrrr}
\tablenum{1}
\tablewidth{0pc}
\tabletypesize{\tiny}
\rotate
\tablecaption{ROSAT PSPC Observations of WR Stars in the MCs}
\tablehead{
\multicolumn{1}{c}{WR $\#$} & 
\multicolumn{1}{c}{WR Name} & 
\multicolumn{1}{c}{Obs.\ ID} & 
\multicolumn{1}{c}{$t_{\rm exp}$} & 
\multicolumn{1}{c}{Offset}  & 
\multicolumn{1}{c}{WR $\#$} & 
\multicolumn{1}{c}{WR Name} & 
\multicolumn{1}{c}{Obs.\ ID} & 
\multicolumn{1}{c}{$t_{\rm exp}$} & 
\multicolumn{1}{c}{Offset}  & 
\multicolumn{1}{c}{WR $\#$} & 
\multicolumn{1}{c}{WR Name} & 
\multicolumn{1}{c}{Obs.\ ID} & 
\multicolumn{1}{c}{$t_{\rm exp}$} & 
\multicolumn{1}{c}{Offset}  \\
\multicolumn{1}{c}{} & 
\multicolumn{1}{c}{} & 
\multicolumn{1}{c}{} & 
\multicolumn{1}{c}{(ks)} & 
\multicolumn{1}{c}{(arcmin)} & 
\multicolumn{1}{c}{} & 
\multicolumn{1}{c}{} & 
\multicolumn{1}{c}{} & 
\multicolumn{1}{c}{(ks)} & 
\multicolumn{1}{c}{(arcmin)} & 
\multicolumn{1}{c}{} & 
\multicolumn{1}{c}{} & 
\multicolumn{1}{c}{} & 
\multicolumn{1}{c}{(ks)} & 
\multicolumn{1}{c}{(arcmin)} 
}

\startdata 
 LMC-WR\,3   & Brey\,3             & rp900320 &  31.3 & 23 &             &                     & rp600099 &   8.8 & 37 & LMC-WR\,54  & Brey\,44a           & rp180251 &  20.2 & 23 \\
 LMC-WR\,5   & Brey\,4             & rp500258 &  13.8 & 34 & LMC-WR\,31  & Brey\,25            & rp400154 &   6.5 &  2 &             &                     & rp180179 &  15.9 & 23 \\
 LMC-WR\,6   & Brey\,5             & rp500263 &  12.7 &  2 &             &                     & rp500093 &   8.7 &  4 &             &                     & rp500100 &  26.6 & 23 \\
 LMC-WR\,7   & Brey\,6             & rp500263 &  12.7 & 21 & LMC-WR\,32  & Brey\,26            & rp400263 &  24.1 & 24 &             &                     & rp500138 &  31.6 & 31 \\
 LMC-WR\,9   & Brey\,7             & rp600098 &  10.5 & 25 & LMC-WR\,33  & Sk\,$-68$\arcdeg73  & rp400154 &   6.5 &  5 &             &                     & rp500140 &  40.0 & 23 \\
             &                     & rp600577 &   9.0 & 25 &             &                     & rp500093 &   8.7 &  8 &             &                     & rp500303 &   9.4 & 23 \\
             &                     & rp900320 &  31.3 & 11 & LMC-WR\,34  & Brey\,28            & rp400263 &  24.1 & 39 &             &                     & rp600100 &  22.7 & 23 \\
 LMC-WR\,10  & Brey\,9             & rp600098 &  10.5 & 21 & LMC-WR\,35  & Brey\,27            & rp500062 &   5.9 & 14 & LMC-WR\,55  & Sk\,$-69$\arcdeg175 & rp180251 &  20.2 & 24 \\
\enddata		           
\end{deluxetable}	           
}			           
			           
\clearpage		           
			           
{\tiny 			           
\begin{deluxetable}{llrrr|llrrr|llrrr}
\tablenum{2}		           
\tablewidth{0pc}	           
\tabletypesize{\tiny}
\rotate
\tablecaption{ROSAT HRI Observations of WR Stars in the MCs}
\tablehead{		    
\multicolumn{1}{c}{WR $\#$} & 
\multicolumn{1}{c}{WR Name} & 
\multicolumn{1}{c}{Obs.\ ID} & 
\multicolumn{1}{c}{$t_{\rm exp}$} & 
\multicolumn{1}{c}{Offset}  & 
\multicolumn{1}{c}{WR $\#$} & 
\multicolumn{1}{c}{WR Name} & 
\multicolumn{1}{c}{Obs.\ ID} & 
\multicolumn{1}{c}{$t_{\rm exp}$} & 
\multicolumn{1}{c}{Offset}  & 
\multicolumn{1}{c}{WR $\#$} & 
\multicolumn{1}{c}{WR Name} & 
\multicolumn{1}{c}{Obs.\ ID} & 
\multicolumn{1}{c}{$t_{\rm exp}$} & 
\multicolumn{1}{c}{Offset}  \\
\multicolumn{1}{c}{} & 
\multicolumn{1}{c}{} & 
\multicolumn{1}{c}{} & 
\multicolumn{1}{c}{(ks)} & 
\multicolumn{1}{c}{(arcmin)} & 
\multicolumn{1}{c}{} & 
\multicolumn{1}{c}{} & 
\multicolumn{1}{c}{} & 
\multicolumn{1}{c}{(ks)} & 
\multicolumn{1}{c}{(arcmin)} & 
\multicolumn{1}{c}{} & 	           
\multicolumn{1}{c}{} & 	           
\multicolumn{1}{c}{} & 	           
\multicolumn{1}{c}{(ks)} &         
\multicolumn{1}{c}{(arcmin)}        
}

\startdata 
 LMC-WR\,3   & Brey\,3            & rh400654 &   2.6 &  7 & LMC-WR\,38  & Brey\,31           & rh400353 &   7.9 & 13 &             &                    & rh400125 &   4.9 & 11 \\
 LMC-WR\,9   & Brey\,7            & rh900321 &  32.2 & 10 &             &                    & rh400355 &   4.5 & 16 &             &                    & rh400658 &   5.9 & 11 \\
 LMC-WR\,10  & Brey\,9            & rh900321 &  32.2 &  0 & LMC-WR\,39  & Brey\,32           & rh400356 &   3.1 & 17 &             &                    & rh500470 &  28.7 & 17 \\
 LMC-WR\,13  & Sk$-66$\arcdeg40   & rh900321 &  32.2 &  8 &             &                    & rh600646 &  23.7 &  9 &             &                    & rh500546 &  22.0 & 17 \\
 LMC-WR\,14  & Brey\,11           & rh400354 &   3.9 & 16 & LMC-WR\,40  & Brey\,33           & rh400356 &   3.1 & 17 &             &                    & rh600775 &  19.7 & 14 \\
 LMC-WR\,19  & Brey\,16           & rh400239 &   3.1 &  9 &             &                    & rh600646 &  23.7 &  9 &             &                    & rh600776 &  20.8 &  5 \\
             &                    & rh500174 &  11.3 & 10 & LMC-WR\,41  & Brey\,35           & rh600033 &   1.7 &  6 & LMC-WR\,61  & Brey\,50           & rh110290 &   1.7 & 12 \\
 LMC-WR\,20  & Brey\,16a          & rh400239 &   3.1 &  8 &             &                    & rh600641 &  26.1 & 18 &             &                    & rh110291 &   1.7 & 12 \\
\enddata
\end{deluxetable}
}

\clearpage

{\tiny
\begin{deluxetable}{llcrrrr}
\tablenum{3}
\tablewidth{0pc}
\tabletypesize{\scriptsize} 
\tablecaption{WR Stars in the MCs Detected by ROSAT Observations}
\tablehead{
\multicolumn{1}{c}{WR $\#$} & 
\multicolumn{1}{c}{WR Name} & 
\multicolumn{1}{c}{Instrument} & 
\multicolumn{1}{c}{$t_{\rm exp}$} & 
\multicolumn{1}{c}{Count Rate} & 
\multicolumn{1}{c}{Counts} & 
\multicolumn{1}{c}{$L_{\rm 0.5-7.0 keV}$} \\
\multicolumn{1}{c}{} &
\multicolumn{1}{c}{} &
\multicolumn{1}{c}{} &
\multicolumn{1}{c}{(ks)} &
\multicolumn{1}{c}{(cnts~s$^{-1}$)} &
\multicolumn{1}{c}{(cnts)} & 
\multicolumn{1}{c}{(ergs s$^{-1}$)} 
}

\startdata
LMC-WR\,10  & Brey\,9   & PSPC & 50.8 & 1.2$\times$10$^{-3}$~ & 61$\pm$10 & (6.7$\pm$1.1)$\times$10$^{33}$ \\
LMC-WR\,38  & Brey\,31  & PSPC & 23.1 & 9.8$\times$10$^{-4}$~ & 23$\pm$6~ & (5.4$\pm$1.4)$\times$10$^{33}$ \\ 
            &           & HRI  & 12.4 & 2.2$\times$10$^{-3}$~ & 28$\pm$8~ & (3.2$\pm$0.9)$\times$10$^{34}$ \\
LMC-WR\,39  & Brey\,32  & PSPC & 31.6 & 1.0$\times$10$^{-3}$~ & 31$\pm$9~ & (5.5$\pm$1.8)$\times$10$^{33}$ \\
LMC-WR\,42  & Brey\,34  & PSPC & 31.6 & 1.2$\times$10$^{-3}$~ & 37$\pm$9~ & (6.7$\pm$1.1)$\times$10$^{33}$ \\
            &           & HRI  & 26.8 & 6.0$\times$10$^{-4}$~ & 16$\pm$8~ & (8.6$\pm$4.3)$\times$10$^{33}$ \\
LMC-WR\,47  & Brey\,39  & PSPC & 31.6 & 7.6$\times$10$^{-4}$~ & 24$\pm$9~ & (4.2$\pm$1.6)$\times$10$^{33}$ \\
LMC-WR\,101-103 & R\,140a,b&HRI&134.8 & 2.9$\times$10$^{-3}$~ &390$\pm$30 & (1.6$\pm$0.1)$\times$10$^{35}$ \\
LMC-WR\,116 & Brey\,84  & HRI  &143.5 & 4.7$\times$10$^{-3}$~ &670$\pm$30 & (1.7$\pm$0.1)$\times$10$^{35}$ \\
\enddata
\end{deluxetable}
}

\clearpage

{\tiny
\begin{deluxetable}{lclc|clclclc}
\tablenum{4}
\tablewidth{57.2pc}
\tabletypesize{\tiny} 
\rotate
\tablecaption{Correlation of X-ray Detected WR Stars in the MCs with Other ROSAT Catalogs}
\tablehead{
\multicolumn{1}{c}{WR $\#$} & 
\multicolumn{1}{c}{This Paper} & 
\multicolumn{2}{c}{\underline{~~~~~~~~~~~~~~~~~~1RXH~~~~~~~~~~~~~~~~~~}} & 
\multicolumn{1}{c}{This Paper} & 
\multicolumn{2}{c}{\underline{~~~~~~~~~~~~~~~~~~2RXP~~~~~~~~~~~~~~~~~~}} &
\multicolumn{2}{c}{\underline{~~~~~~~~~~~~~~~~~~1WGA~~~~~~~~~~~~~~~~~~}} &
\multicolumn{2}{c}{\underline{~~~~~~~~~~~~~~~~~~1RXS~~~~~~~~~~~~~~~~~~}} \\
\multicolumn{1}{c}{} &
\multicolumn{1}{c}{HRI Count Rate} & 
\multicolumn{1}{l}{Source Name} & 
\multicolumn{1}{c}{Count Rate} & 
\multicolumn{1}{c}{PSPC Count Rate} & 
\multicolumn{1}{l}{Source Name} & 
\multicolumn{1}{c}{Count Rate} & 
\multicolumn{1}{l}{Source Name} & 
\multicolumn{1}{c}{Count Rate} & 
\multicolumn{1}{l}{Source Name} & 
\multicolumn{1}{c}{Count Rate} \\
\multicolumn{1}{c}{} &
\multicolumn{1}{c}{(cnts~s$^{-1}$)} &
\multicolumn{1}{c}{} &
\multicolumn{1}{c}{(cnts~s$^{-1}$)} &
\multicolumn{1}{c}{(cnts~s$^{-1}$)} &
\multicolumn{1}{c}{} &
\multicolumn{1}{c}{(cnts~s$^{-1}$)} &
\multicolumn{1}{c}{} &
\multicolumn{1}{c}{(cnts~s$^{-1}$)} &
\multicolumn{1}{c}{} &
\multicolumn{1}{c}{(cnts~s$^{-1}$)} 
}

\startdata
LMC-WR\,10      &              \nodata & J045633.1--662828 & 8.5$\times$10$^{-4}$ & 1.2$\times$10$^{-3}$ & J045635.7--662815 & 9.7$\times$10$^{-3}$ & J0456.5--6628 & 1.6$\times$10$^{-3}$ & 045635.6--662819 & 2.8$\times$10$^{-2}$  \\
LMC-WR\,38      & 2.2$\times$10$^{-3}$ & J052605.4--672958 & 2.9$\times$10$^{-3}$ & 9.8$\times$10$^{-4}$ & J052604.8--673002 & 3.8$\times$10$^{-3}$ & J0526.0--6730 & 1.8$\times$10$^{-3}$ &          \nodata &              \nodata  \\
LMC-WR\,39      &              \nodata &           \nodata &              \nodata & 1.0$\times$10$^{-3}$ &           \nodata &              \nodata & J0526.4--6850 & 2.4$\times$10$^{-3}$ &          \nodata &              \nodata  \\
LMC-WR\,42      & 6.0$\times$10$^{-4}$ &           \nodata &              \nodata & 1.2$\times$10$^{-3}$ & J052643.4--684950 & 3.3$\times$10$^{-3}$ &       \nodata &              \nodata &          \nodata &              \nodata  \\
LMC-WR\,101-103 & 2.9$\times$10$^{-3}$ &  & 5.8$\times$10$^{-3}$ &              \nodata &           \nodata &              \nodata &  & 1.9$\times$10$^{-2}$ &        \nodata &            \nodata  \\
LMC-WR\,116     & 4.7$\times$10$^{-3}$ &  & 3.9$\times$10$^{-3}$ &              \nodata & J053844.8--690602 & 1.3$\times$10$^{-1}$ &  & 4.6$\times$10$^{-2}$ &        \nodata &              \nodata  \\
\enddata
\end{deluxetable}
}

\clearpage

{\tiny
\begin{deluxetable}{lllrr|lllrr}
\tablenum{5}
\tablewidth{0pc}
\tabletypesize{\tiny} 
\tablecaption{WR Stars in the MCs Undetected by ROSAT Observations}
\tablehead{
\multicolumn{1}{c}{WR $\#$} & 
\multicolumn{1}{c}{WR Name} & 
\multicolumn{1}{c}{Instrument} & 
\multicolumn{1}{c}{$t_{\rm exp}$} & 
\multicolumn{1}{c}{$3\sigma$ Upper Limit} & 
\multicolumn{1}{c}{WR $\#$} & 
\multicolumn{1}{c}{WR Name} & 
\multicolumn{1}{c}{Instrument} & 
\multicolumn{1}{c}{$t_{\rm exp}$} & 
\multicolumn{1}{c}{$3\sigma$ Upper Limit} \\
\multicolumn{1}{c}{} &
\multicolumn{1}{c}{} &
\multicolumn{1}{c}{} &
\multicolumn{1}{c}{(ks)} &
\multicolumn{1}{c}{(cnts~s$^{-1}$)} & 
\multicolumn{1}{c}{} &
\multicolumn{1}{c}{} &
\multicolumn{1}{c}{} &
\multicolumn{1}{c}{(ks)} &
\multicolumn{1}{c}{(cnts~s$^{-1}$)} 
}

\startdata
LMC-WR\,3   & Brey\,3             & PSPC &  31.3 & 6.5$\times$10$^{-4}$~ & LMC-WR\,50  & Brey\,41            & PSPC &  51.8 & 5.5$\times$10$^{-4}$~ \\
            &                     & HRI  &   2.6 & 2.7$\times$10$^{-3}$~ &             &                     & HRI  &  96.7 & 4.6$\times$10$^{-4}$~ \\
LMC-WR\,5   & Brey\,4             & PSPC &  13.8 & 1.4$\times$10$^{-3}$~ & LMC-WR\,51  & Brey\,42            & PSPC &  31.6 & 7.4$\times$10$^{-4}$~ \\
LMC-WR\,6   & Brey\,5             & PSPC &  12.7 & 1.2$\times$10$^{-3}$~ &             &                     & HRI  &  50.6 & 7.1$\times$10$^{-4}$~ \\
LMC-WR\,7   & Brey\,6             & PSPC &  12.7 & 9.5$\times$10$^{-4}$~ & LMC-WR\,52  & Brey\,43            & PSPC &  10.0 & 1.0$\times$10$^{-3}$~ \\
LMC-WR\,9   & Brey\,7             & PSPC &  50.8 & 4.2$\times$10$^{-4}$~ &             &                     & HRI  &   4.6 & 3.9$\times$10$^{-3}$~ \\
            &                     & HRI  &  32.2 & 6.5$\times$10$^{-4}$~ & LMC-WR\,53  & Brey\,44            & PSPC &  10.9 & 1.2$\times$10$^{-3}$~ \\
LMC-WR\,10  & Brey\,9             & HRI  &  32.2 & 7.3$\times$10$^{-4}$~ &             &                     & HRI  &  48.7 & 6.2$\times$10$^{-4}$~ \\
\enddata
\end{deluxetable}
}

\clearpage

\noindent
Figure~1. 

\noindent
\emph{ROSAT} HRI ({\it left}) and PSPC ({\it center}) smoothed X-ray and 
DSS optical ({\it right}) images of the WR stars in the MCs with detected 
X-ray emission.  
The \emph{ROSAT} HRI and PSPC images are overlaid with their corresponding 
X-ray contours, while the optical DSS images are overlaid with the
PSPC X-ray contours.  
The X-ray contour levels are 3$\sigma$, 6$\sigma$, 9$\sigma$, 12$\sigma$, 
15$\sigma$, 25$\sigma$, 35$\sigma$, ... above the background level.  
The positions of WR stars in the LMC given by \citet{BAT99} are 
marked with a '+' sign.  \\

\noindent
Figure~2. 

\noindent
\emph{ROSAT} HRI smoothed X-ray ({\it left}) and \emph{HST} optical 
({\it right}) images of R\,136 ({\it top}) and R\,140 ({\it bottom}).  
As in Fig.~1, the \emph{ROSAT} HRI and \emph{HST} images are overlaid with 
X-ray contours at 3$\sigma$, 6$\sigma$, 9$\sigma$, 12$\sigma$, 15$\sigma$, 
25$\sigma$, 35$\sigma$, ... above the background level, and the ``+'' sign 
marks the locations of WR stars in the LMC given by \citet{BAT99}.  

\end{document}